\documentclass[aps,prb,twocolumn,groupedaddress,draft,intlimits,amsmath,amssymb,floats]{revtex4}

\usepackage{bm}
\usepackage[final]{graphicx}
\usepackage{epsfig}
\usepackage{subfigure}

\begin{document}

\title{Doping evolution of spin and charge excitations in the Hubbard model} 

\author{Y. F. Kung$^{1,2}$}
\author{E. A. Nowadnick$^{1,2,3}$}
\author{C. J. Jia$^{2,4}$}
\author{S. Johnston$^{5,6}$}
\author{B. Moritz$^{2,7}$}
\author{R. T. Scalettar$^8$}
\author{T. P. Devereaux$^{2,9}$}

\affiliation{$^1$Department of Physics, Stanford University, Stanford, CA 94305, USA}
\affiliation{$^2$Stanford Institute for Materials and Energy Sciences, 
SLAC National Accelerator Laboratory and Stanford University, Menlo Park, California 94025, USA}
\affiliation{$^3$School of Applied and Engineering Physics, Cornell University, Ithaca, NY 14853 USA}
\affiliation{$^4$Department  of Applied Physics,  Stanford  University,  California 94305,  USA}
\affiliation{$^5$Department of Physics and Astronomy, University of Tennessee, Knoxville, TN 37996, USA}
\affiliation{$^6$Joint Institute for Advanced Materials, The University of Tennessee, Knoxville, TN 37996, USA}
\affiliation{$^7$Department of Physics and Astrophysics, University of North Dakota, Grand Forks, ND 58202, USA}
\affiliation{$^8$Department of Physics, University of California - Davis, CA 95616, USA}
\affiliation{$^9$Geballe Laboratory for Advanced Materials, Stanford University, Stanford, CA 94305, USA}

\begin{abstract}
To shed light on how electronic correlations vary across the phase diagram of the cuprate superconductors, we examine the doping evolution of spin and charge excitations in the single-band Hubbard model using determinant quantum Monte Carlo (DQMC).  In the single-particle response, we observe that the effects of correlations weaken rapidly with doping, such that one may expect the random phase approximation (RPA) to provide an adequate description of the two-particle response.  In contrast, when compared to RPA, we find that significant residual correlations in the two-particle excitations persist up to $40\%$ hole and $15\%$ electron doping (the range of dopings achieved in the cuprates).  These fundamental differences between the doping evolution of single- and multi-particle renormalizations show that conclusions drawn from single-particle processes cannot necessarily be applied to multi-particle excitations.  Eventually, the system smoothly transitions via a momentum-dependent crossover into a weakly 
correlated metallic state where the spin and charge excitation spectra exhibit similar behavior and where RPA provides an adequate description. 
\end{abstract}


\maketitle

\section{Introduction}
A full understanding of the cuprate phase diagram has been prevented in part by difficulties in obtaining well-controlled theories for the doping evolution of electronic excitations.  This is due to the lack of exact solutions to the Hubbard model in two dimensions, the standard model thought to contain the low-energy physics of the cuprates that captures aspects of magnetic properties seen in experiments \cite{Dagotto_RMP_1994}.  At low doping, strong coupling treatments of the large-$U$ Hubbard model describe the dispersion and intensity of magnon excitations observed via neutron scattering \cite{Scalapino_RMP_2012}.  At high doping, it is believed that spin excitations behave like weakly interacting particle-hole excitations governed by the underlying free particle kinetic energy, with a weak influence from the Hubbard $U$.  If correct, this limit can be adequately represented by the random phase approximation (RPA) as a proxy for more exact treatments \cite{Chen_PRL_1991}.  Many studies have assumed weak 
correlations in doped cuprates, so RPA has been used to address spin and charge excitations observed by neutron and Raman scattering, as well as the formation of a $d$-wave superconducting ground state  \cite{Devereaux_RMP_2007,Bulut_PRB_1993,Monthoux_PRB_1994,Scalapino_PhysRep_1995,Dahm_SSC_1997,Maier_PRB_2007_1,Maier_PRB_2007_2,Scalapino_RMP_2012}.

However, a set of surprising results have emerged from recent resonant inelastic x-ray scattering measurements (RIXS) in a variety of cuprates \cite{Ament_RMP_2011}.  In the hole-doped La$_{2-x}$Sr$_x$CuO$_4$, YBa$_2$Cu$_3$O$_{6+x}$, Y$_{1-x}$Ca$_x$Ba$_2$Cu$_3$O$_{6+x}$, Tl$_2$Ba$_2$CuO$_{6+x}$, and Bi$_2$Sr$_2$CaCu$_2$O$_{8+x}$ families \cite{LeTacon_NatPhys_2011,LeTacon_PRB_2013,Dean_PRL_2013,Dean_NatMat_2013,Lee_NatPhys_2014}, high-energy magnons or paramagnons on the antiferromagnetic zone boundary (AFZB) persist from the parent compounds into the heavily overdoped regime, showing little doping dependence up to $40\%$ hole doping where the system is believed to exhibit Fermi-liquid-like behavior in the single-particle response or transport.  In the electron-doped Nd$_{2-x}$Ce$_x$CuO$_4$ family \cite{Ishii_PRL_2005,Li_PRB_2008}, momentum-dependent low-energy charge excitations have been found over a large energy range and surprisingly, the magnetic excitations harden to high energies when doped beyond the 
antiferromagnetic phase.  Both computational and analytical techniques have been brought to bear on these results.  Exact diagonalization and determinant quantum Monte Carlo (DQMC) have captured the momentum and doping dependence of the AFZB paramagnons \cite{Jia_NatComm_2014}.  On the other hand, RPA has been used to conclude that the collective spin and charge excitations have similar low-energy behavior \cite{Guarise_NatComm_2014}.  Beyond the spin response, the charge excitations may be associated with charge ordering observed in the cuprate pseudogap regime, so it is useful to systematically explore how both spin and charge excitations evolve with doping throughout the Brillouin zone, and to what degree the response functions can be approximated using RPA.

In this study, we compute spin and charge susceptibilities of the single-band Hubbard model \cite{Anderson_PR_1959,Hubbard_PRSLA_1963,Dagotto_RMP_1994} throughout the first Brillouin zone (Fig. 1) via DQMC, a numerically exact imaginary-time auxiliary-field technique \cite{BSS_PRD_1981,Hirsch_PRB_1985,White_PRB_1989}.  The susceptibilities are compared to those calculated with RPA \cite{Berk_PRL_1966}, a formalism originally developed for weakly interacting systems that is expected to become an increasingly good approximation as the doping level increases.  Our calculations reveal that the influence of correlations on the multi-particle excitations persists to higher dopings than suggested by quantities related to the single-particle response \cite{Hirsch_PRB_1985}.  At even higher doping, RPA provides an adequate description of the response functions, which show a smooth momentum-dependent crossover into a weakly correlated metallic state.

\begin{figure}[t!]
	\includegraphics[scale=0.6]{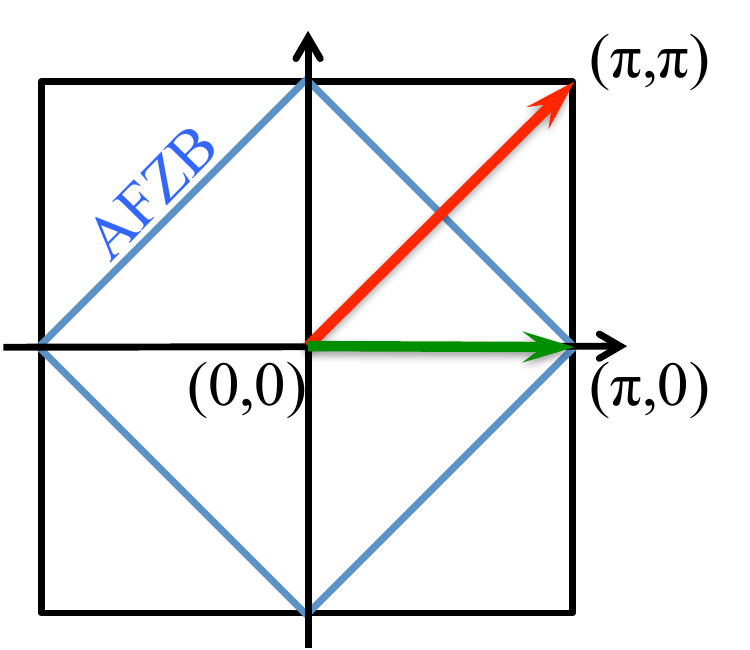}
	\centering
	\caption[Cartoon of the first Brillouin zone.]{Cartoon of the first Brillouin zone showing the antiferromagnetic zone boundary (AFZB) and high-symmetry cuts.  The nodal direction, from zone center to $(\pi,\pi)$, is indicated with a red arrow.  The antinodal direction, from zone center to $(\pi,0)$, is marked with a green arrow.}
\end{figure}

\section{Model and Methods}
The single-band Hubbard Hamiltonian describes strongly correlated electrons on a lattice:
\begin{eqnarray}
H&=&\sum_{\langle i,j \rangle \sigma} t_{ij}(c^\dagger_{i\sigma} c^{\phantom{\dagger}}_{j\sigma} + h.c.)
+\sum_{\{ i,j \} \sigma} t^\prime_{ij}(c^\dagger_{i\sigma} c^{\phantom{\dagger}}_{j\sigma} + h.c.)\nonumber\\
&&-\mu\sum_{i\sigma}n_{i\sigma}+U\sum_i n_{i\uparrow}n_{i\downarrow}
\end{eqnarray}
where $c^\dagger_{i\sigma}$ ($c^{\phantom{\dagger}}_{i\sigma}$) creates (annihilates) a particle with spin $\sigma$ on site $i$, and $n_{i\sigma}=c^\dagger_{i\sigma}c^{\phantom{\dagger}}_{i\sigma}$ is the number operator.  The nearest- and next-nearest-neighbor hoppings are controlled by $t$ and $t^\prime$, respectively, and $U$ is the on-site Coulomb interaction that penalizes double occupancy.  We work with a canonical parameter set, $t^\prime=-0.3t$ and $U=8t$, so the ground state is a strongly correlated Mott insulator in the undoped system \cite{Lee_RMP_2006} and upon hole-doping the system will possess a hole-like Fermi surface.  As usual all energies are expressed in units of $t$.  The chemical potential $\mu$ is adjusted to give between $15\%$ electron and $75\%$ hole doping.  Without particle-hole symmetry, DQMC exhibits a significant sign problem \cite{White_PRB_1989}, so we work at an inverse temperature $\beta=3/t$ to give reasonable statistics.  The imaginary-time spin and charge correlators are computed as
\begin{eqnarray}
\chi_{s,c}(\textbf{q},\tau)=\langle T_\tau \hat{O}^{\phantom{\dagger}}_{s,c}(\textbf{q},\tau) \hat{O}^\dagger_{s,c}(\textbf{q},0) \rangle,
\end{eqnarray}
where $\hat{O}_s = \sum_i e^{i\textbf{q}\cdot \textbf{R}_i} (n_{i \uparrow} - n_{i \downarrow})$, $\hat{O}_c = \sum_i e^{i\textbf{q}\cdot \textbf{R}_i} (n_{i \uparrow} + n_{i \downarrow})$, and $n_{i \sigma}$ is the number operator.  The correlators are analytically continued using the Maximum Entropy method to obtain the real-frequency susceptibilities \cite{Jarrell_MEM_1996}.

\begin{figure}[t!]
	\includegraphics[width=\columnwidth]{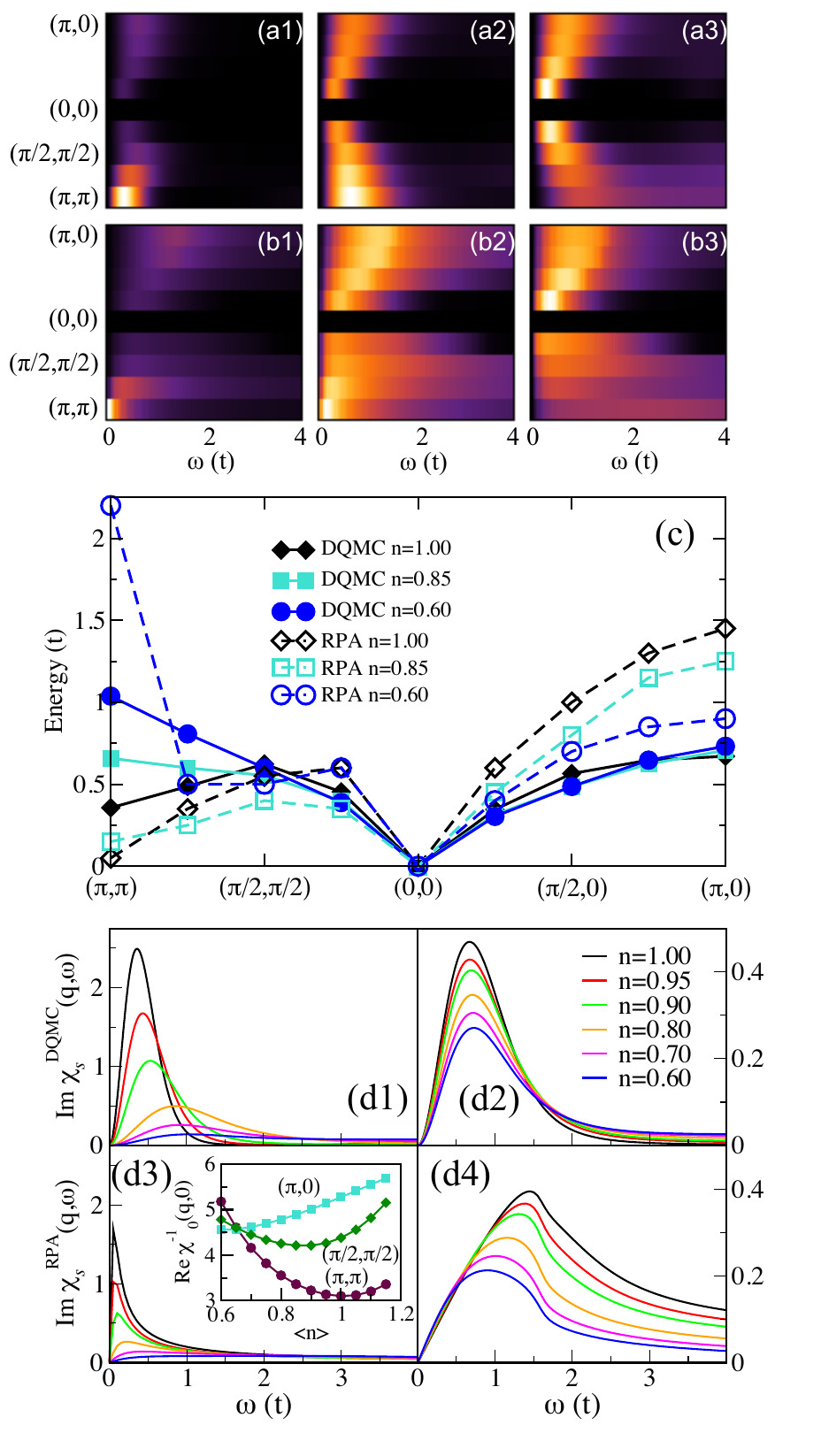}
	\caption{The spin susceptibilities along high-symmetry cuts in the Brillouin zone are calculated using DQMC [panels (a1)-(a3)] and RPA [panels (b1)-(b3)], for three hole dopings [n=1.00 in (a1) and (b1), n=0.85 in (a2) and (b2), and n=0.60 in (a3) and (b3)].  The maximum of the color scale is set by the highest intensity in each panel.  (c) The peak energies in (a1)-(a3) and (b1)-(b3) are plotted versus momentum to highlight the doping trends.  Spin susceptibilities at the representative momentum points $(\pi,\pi)$ and $(\pi,0)$ from DQMC [(d1) and (d2), respectively] and RPA [(d3) and (d4), respectively] are shown for hole doping ranging from $0\%$ to $40\%$.}
\end{figure}

RPA susceptibilities are determined for the same temperature and doping range as DQMC, and are normalized to the DQMC two-particle sum rule for ease of comparison.  The RPA susceptibilities are computed from:
\begin{eqnarray}
\chi_{s,c}^{\mathrm{RPA}}(\mathbf{q},\omega) = \frac{\chi_0(\mathbf{q},\omega)}{1\mp\bar{U}\chi_0(\mathbf{q},\omega)},
\end{eqnarray}
where $-$($+$) corresponds to the spin (charge) susceptibility, $\bar{U}$ is the effective interaction strength, and $\chi_0$ is the non-interacting Lindhard susceptibility \cite{Dahm_SSC_1997,Mahan}
\begin{eqnarray}
\chi_0(\textbf{q},\omega) = \frac{1}{N}\sum_k \frac{f(\epsilon_{k+q})-f(\epsilon_k)}{i\omega - (\epsilon_{k+q}-\epsilon_k)}.
\end{eqnarray}
The bandstructure is $\epsilon_k = -2t(\cos{k_x}+\cos{k_y}) - 4t^\prime \cos{k_x}\cos{k_y}$ and the Fermi function is $f(\epsilon_k) = \{1+\exp{[\beta(\epsilon_k-\mu)]}\}^{-1}$, with the chemical potential $\mu$ controlling the filling.   The inset in Fig. 2(d3) shows $\bar{U}_{\mathrm{max}}= \mathrm{Re}\chi_0^{-1}(q,0)$, the location of the new pole in $\chi_s^{\mathrm{RPA}}$, plotted versus filling for $(\pi,\pi)$, as well as for $(\pi,0)$ and $(\pi/2,\pi/2)$ on the AFZB.  $\bar{U}$ is most tightly constrained by $(\pi,\pi)$.

To isolate doping-dependent bandstructure effects from doping-dependent changes to effective interactions that can be harder to predict, we use a single value of $\bar{U}$ to calculate the RPA susceptibility for all momenta and all doping levels.  Because the low-doping behavior of the spin response in cuprates, and the single-band Hubbard model, is well known, we set $\bar{U}=3t$ such that the RPA response mimics the spin response near the AF instability at $(\pi,\pi)$ for $0\%$ doping which captures the experimental neutron scattering results \cite{Headings_PRL_2010} and matches the DQMC trends by construction.   These low-energy spin excitations near $(\pi,\pi)$ also may be integral to pairing in a broad class of superconductors  \cite{Scalapino_RMP_2012}, and their doping evolution will thus be well captured by both computational methods.  We note that this value for $\bar{U}$ falls within the range of values obtained in previous studies \cite{Bulut_PRB_1993,Maier_PRB_2007_1,Maier_PRB_2007_2}; however, one should note again that for the chosen set of parameters the undoped, single-band Hubbard model possesses a Mott insulating ground state, and should not be viewed as weakly correlated. The expression for $\chi_c^{\mathrm{RPA}}$ does not contain a new pole, so its behavior is primarily determined by that of $\chi_0$ and thus set by the non-interacting bandstructure.

\section{Doping evolution of spin and charge susceptibilities}
Figure 2 summarizes the hole doping evolution of the DQMC [(a1)-(a3)] and RPA [(b1)-(b3)] spin susceptibilities along the nodal (from $\mathbf{q}=(0,0)$ to $\mathbf{q}=(\pi,\pi)$) and antinodal (from $\mathbf{q}=(0,0)$ to $\mathbf{q}=(\pi,0)$) cuts in the first Brillouin zone.  For momenta near $(\pi,\pi)$, doping both reduces and shifts spectral weight to higher energies, with a good comparison between DQMC and RPA as doping increases, a mirror of the results from neutron scattering experiments \cite{Wakimoto_PRL_2007,Fujita_JPSJ_1012} and a by-product of the choice of $\bar{U}$.  On the other hand, along the antinodal direction the peak in the spin response remains unchanged with doping in the DQMC calculations but softens considerably in RPA, highlighted in Fig. 2(c).  The strong disagreement between DQMC and RPA in large portions of the Brillouin zone as doping increases indicates that the influence of correlations on the spin response remains considerable to relatively large doping levels.

\begin{figure}[t!]
	\includegraphics[width=\columnwidth]{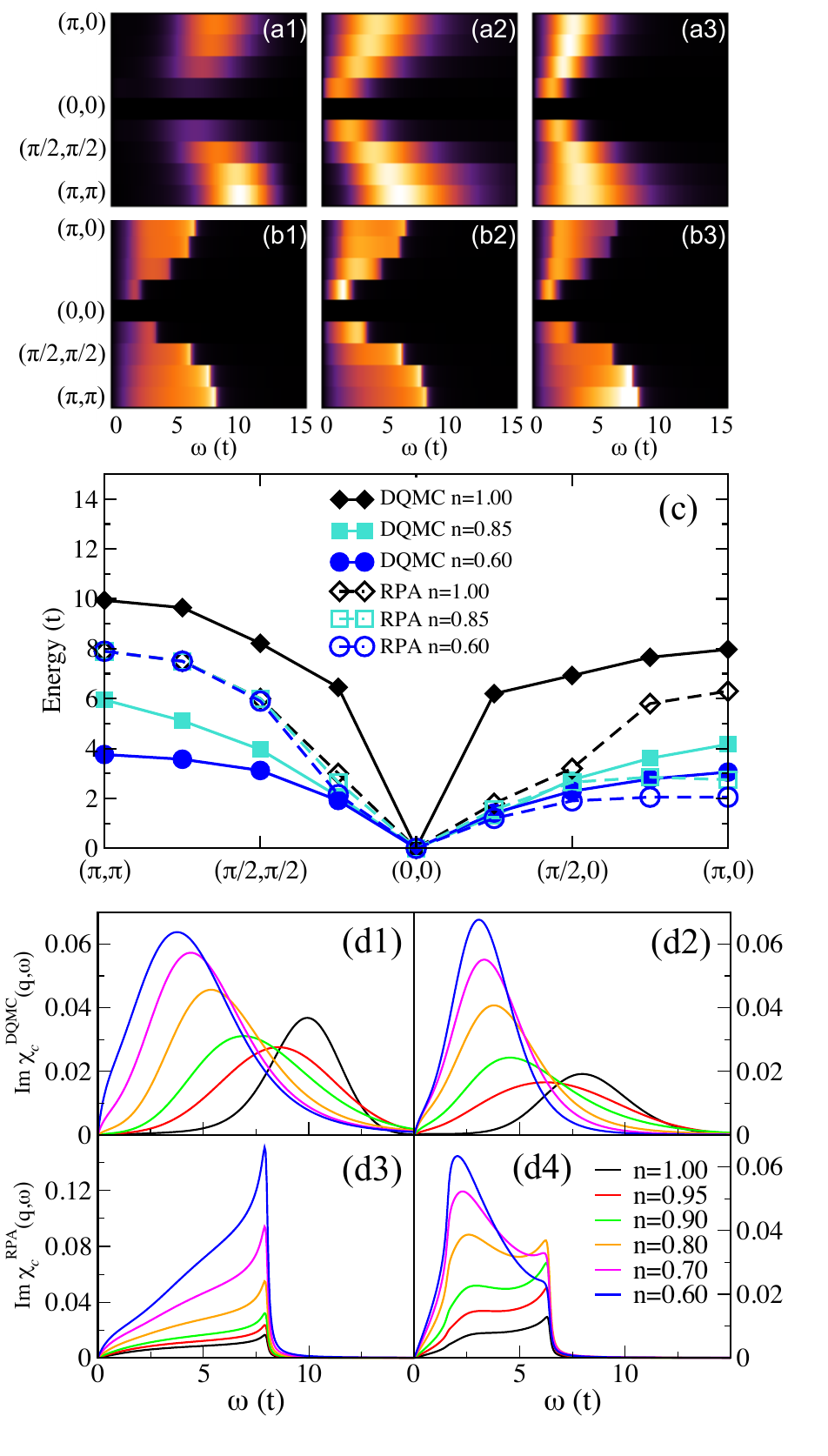}
	\caption{The charge susceptibilities along high-symmetry cuts in the Brillouin zone are calculated using DQMC [panels (a1)-(a3)] and RPA [panels (b1)-(b3)], for three hole dopings [n=1.00 in (a1) and (b1), n=0.85 in (a2) and (b2), and n=0.60 in (a3) and (b3)].  The maximum of the color scale is set by the highest intensity in each panel.  (c) The peak energies in (a1)-(a3) and (b1)-(b3) are plotted versus momentum to highlight the doping trends.  Charge susceptibilities at the representative momentum points $(\pi,\pi)$ and $(\pi,0)$ from DQMC [(d1) and (d2), respectively] and RPA [(d3) and (d4), respectively] are shown for hole doping ranging from $0\%$ to $40\%$.}
\end{figure}

Since $(\pi,\pi)$ and $(\pi,0)$ are representative of the behavior along the nodal and antinodal directions, respectively, Fig. 2(d1)-(d4) focuses on the evolution from $0\%$ to $40\%$ hole doping at those momenta.  At $0\%$ doping, the $(\pi,\pi)$ spin response peaks strongly at low energy, but it broadens and hardens with increasing doping, as determined by the non-interacting bandwidth.  These findings agree with previous DQMC and dynamical cluster approximation calculations \cite{Preuss_PRL_1997,Groeber_PRB_2000,Hochkeppel_PRB_2008}.  Due to proximity to the pole in Eq. 3, the low-doping RPA spin excitations at $(\pi,\pi)$ generally occur at lower energies than in DQMC and their peaks are much sharper.  In addition, the RPA response has longer tails set by the full non-interacting bandwidth.  At $(\pi,0)$, the DQMC and RPA spin responses remain quite different up to at least $40\%$ hole doping.  The RPA spin susceptibilities soften significantly with doping, whereas the DQMC spin response shows persistent excitations in agreement with experiments.

Figure 3 summarizes the hole doping evolution of the charge susceptibilities as calculated using DQMC [(a1)-(a3)] and RPA [(b1)-(b3)] throughout the first Brillouin zone.  Along both the nodal and antinodal cuts, the spectral weight of the response evaluated using DQMC is located at high energies (determined at $0\%$ doping by the Hubbard $U$) and the spectra show a charge gap that decreases with doping \cite{Preuss_PRL_1997,Groeber_PRB_2000}.  On the other hand, the RPA susceptibility shows no charge gap and is dominated by the peak in the Lindhard susceptibility at the band edge.  With doping, the lineshape of the RPA charge response along the nodal cut changes little [Figs. 3(b1)-(b3)], although spectral weight along the antinodal direction increases at lower energies in agreement with the response from DQMC.  Figure 3(c) highlights the doping evolution of the peak positions and Figs. 3(d1)-(d4) focus on a more detailed doping dependence for representative momenta in the nodal, $(\pi,\pi)$, and antinodal, $(\pi,0)$, 
directions.

\begin{figure}[t!]
	\includegraphics[width=\columnwidth]{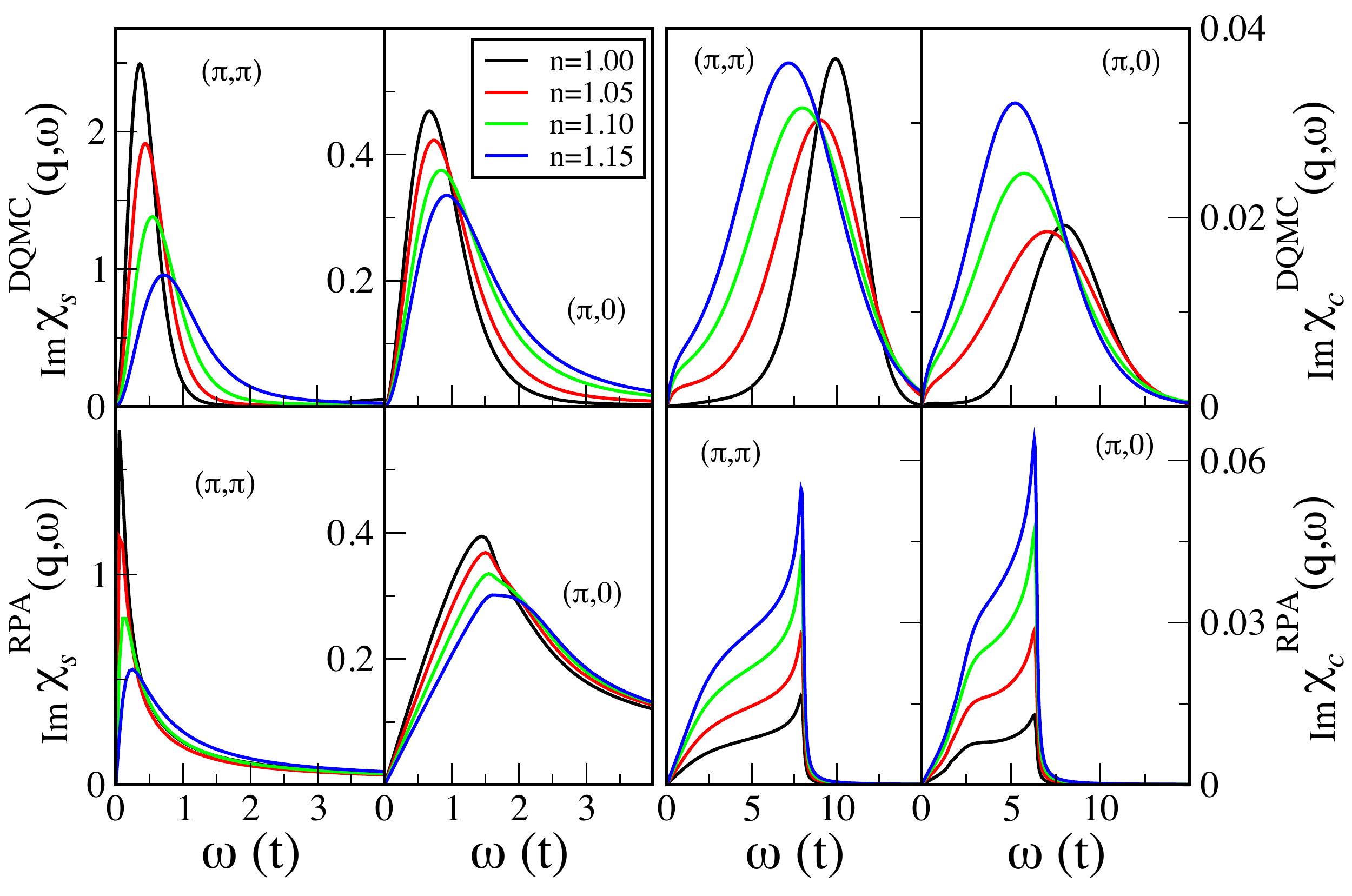}
	\caption{The spin and charge susceptibilities are calculated by DQMC and RPA for different levels of electron doping.  While the DQMC spin susceptibility shows the same doping trend as RPA at $(\pi,\pi)$ and $(\pi,0)$, suggesting decreasing correlations, the RPA charge susceptibility is dominated by the band edge, unlike DQMC.}
\end{figure}

Upon electron doping, the DQMC and RPA spin susceptibilities (Fig. 4) show similar behavior, hardening and broadening with increasing doping at $(\pi,\pi)$, again by construction.  At $(\pi,0)$, both $\mathrm{Im}\chi_s^{\mathrm{RPA}}$ and $\mathrm{Im}\chi_s^{\mathrm{DQMC}}$ harden and decrease in intensity with doping.  The spin response thus exhibits an electron-hole doping asymmetry, as the doping trend near $(\pi,0)$ is different from that on the hole-doped side (Fig. 2).  This suggests that correlation effects on the spin susceptibilities become less relevant more rapidly with electron than hole doping.  In the charge channel, the DQMC and RPA responses exhibit different doping trends similar to the evolution on the hole-doped side:  the charge gap along the nodal and antinodal directions at $0\%$ doping in the DQMC charge susceptibility vanishes with electron doping whereas the RPA charge susceptibility shows no gap, and the low-energy structure in $\mathrm{Im} \chi_c^{\mathrm{RPA}}$ remains essentially unchanged up to $15\%$ electron doping for both $(\pi,\pi)$ and $(\pi,0)$, again manifesting the system's electron-hole doping asymmetry.

\begin{figure}[t!]
	\includegraphics[width=\columnwidth]{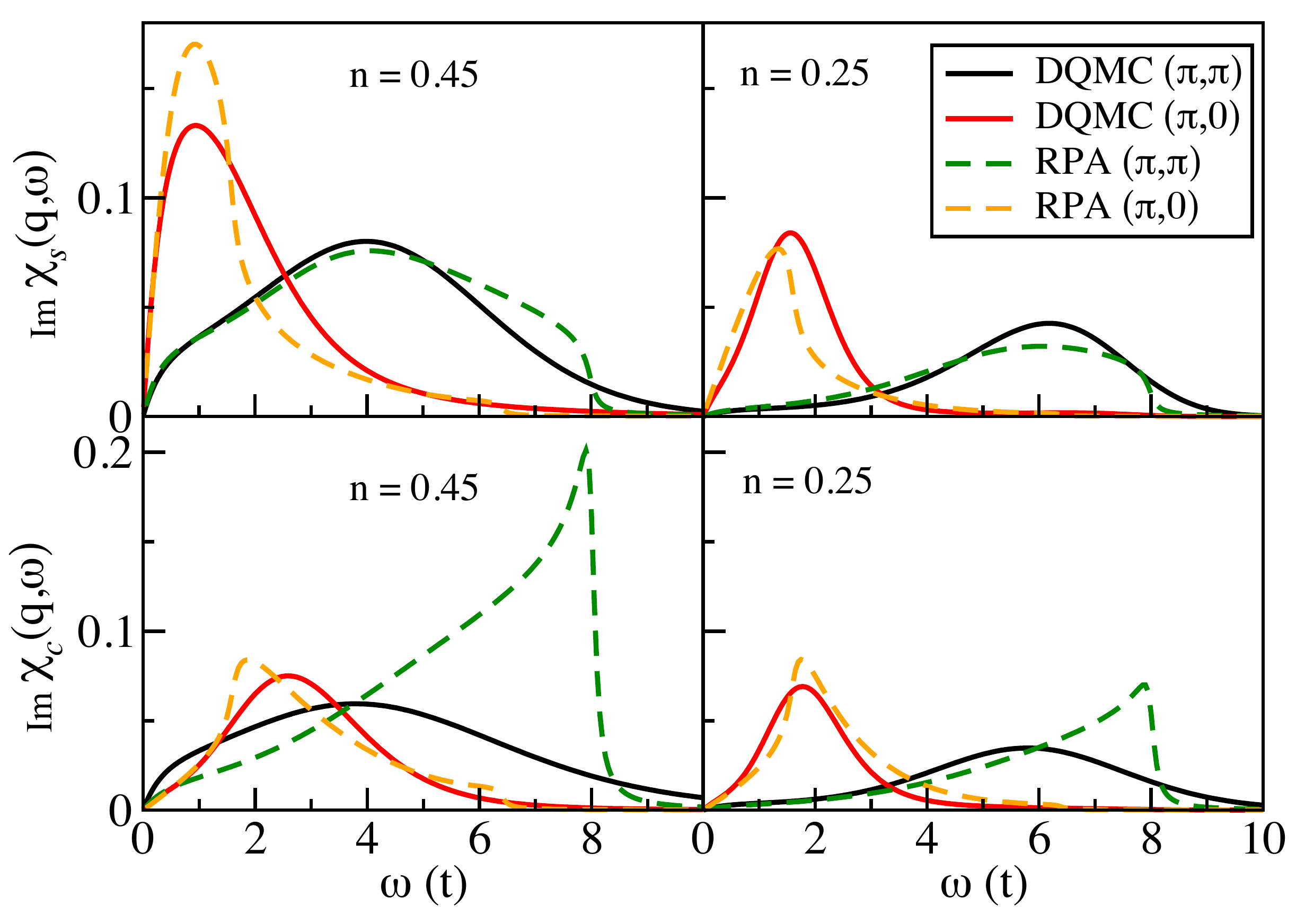}
	\caption{The spin and charge susceptibilities calculated by DQMC and RPA are shown for representative momenta at $55\%$ and $75\%$ hole doping to illustrate the gradual weakening of correlations at extremely high doping.}
\end{figure}

As the system is doped to extremely high levels, the correlations continue to weaken until the system smoothly transitions into a weakly correlated metal, where RPA provides a fairly adequate description of the spin and charge responses across the Brillouin zone.  As shown in Fig. 5, at $55\%$ and even more so at $75\%$ hole doping, $\mathrm{Im} \chi_s^{\mathrm{DQMC}}$ and $\mathrm{Im}\chi_s^{\mathrm{RPA}}$ agree both qualitatively, and even to some degree quantitatively, at low energies.  In addition, the DQMC spin peak broadens significantly with increasing doping until it closely resembles the charge susceptibility at $75\%$ hole doping, indicating that the response in the spin channel essentially can be viewed as a trivial spin flip on top of the charge excitations.  Figure 5 thus demonstrates that the influence of correlations on the multi-particle response can decrease significantly, but only at doping levels well beyond those where weak coupling approaches typically already have been applied for the cuprates.    

\section{Correlations in single- versus two-particle correlators}
An important subtlety in examining how correlations evolve with doping is that single- and multi-particle quantities exhibit fundamentally different renormalizations.  The spin and charge susceptibilities reveal that correlations in two-particle quantities persist to significantly higher dopings than suggested by single-particle correlators \cite{Hirsch_PRB_1985}, necessitating the study of both to resurrect the full richness of the phase diagram.  Figure 6(a) shows the doping dependence of the compressibility for $t^\prime=0$ and $-0.3t$, with the next-nearest-neighbor hopping breaking particle-hole symmetry.  Because the compressibility is zero in the Mott plateau and finite elsewhere, it could be interpreted as a proxy of how strong the correlations are.  For both $t^\prime=0$ and $-0.3t$, the compressibility vanishes near half filling, corresponding to the Mott plateau, as expected.  However, it recovers rapidly with both hole and electron doping, suggesting that the correlation effects weaken rapidly away from half filling.

On the other hand, the nearest-neighbor equal-time spin-spin correlation function [Fig. 6(b)] suggests that correlations extend to higher doping levels.  For $t^\prime=0$, the spin correlations persist well away from half filling with both hole and electron doping.  Next-nearest-neighbor hopping strongly suppresses the correlation function above $30\%$ hole doping, suggesting that correlations have weakened significantly by that doping level, but the magnitude of the correlation function is actually enhanced slightly on the electron-doped side.  The equal-time spin-spin correlation function thus implies that correlation effects persist to significantly higher doping levels than can be seen in the compressibility.

\begin{figure}[t!]
	\includegraphics[width=\columnwidth]{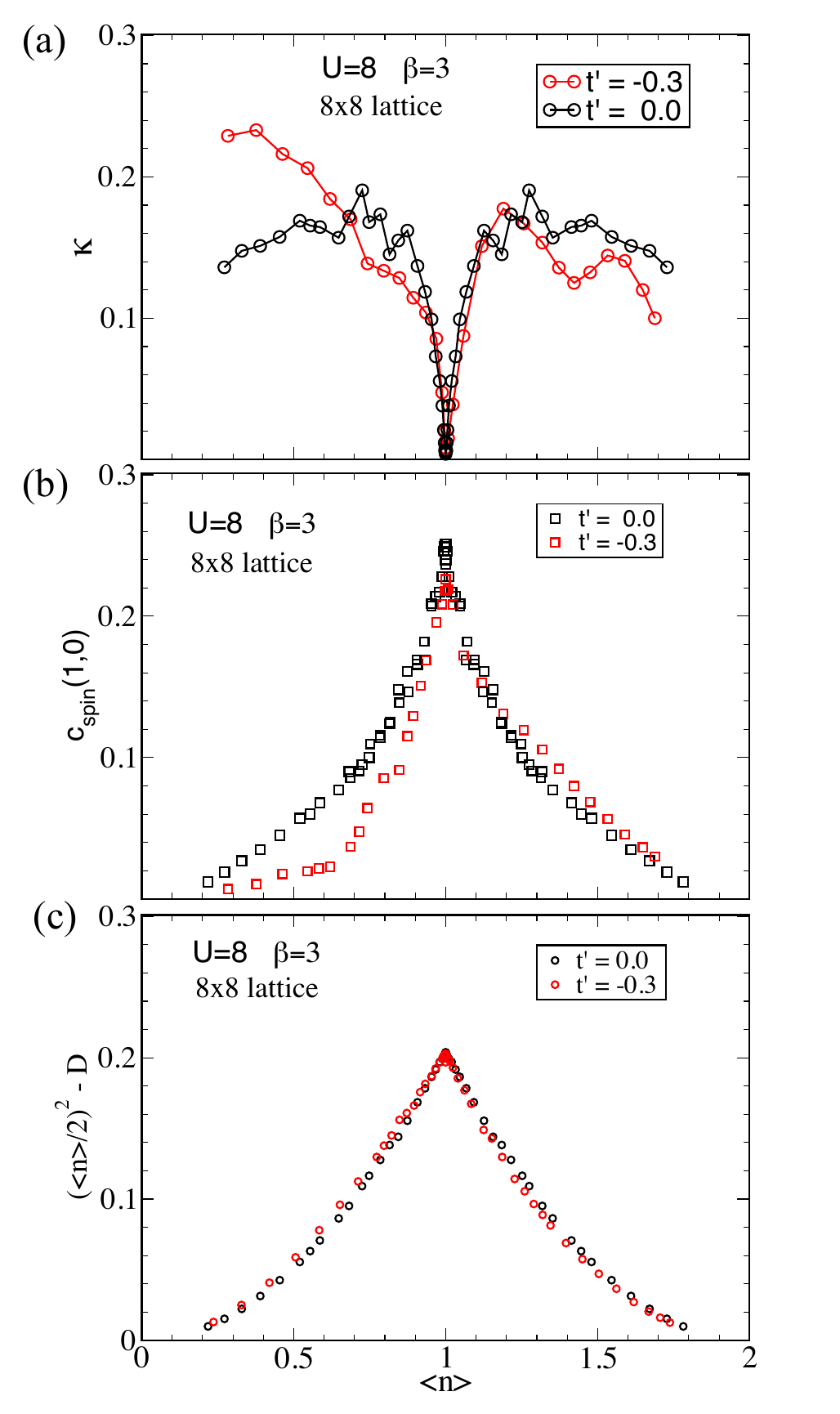}
	\caption{The doping trends of the (a) compressibility, (b) equal-time spin-spin correlation function, and (c) difference between the double occupancy $D=\langle n_\uparrow n_\downarrow \rangle$ from its uncorrelated value of $(\langle n \rangle/2)^2$ are shown for next-nearest-neighbor hopping $t^\prime=0$ and $-0.3t$.}
\end{figure}

The difference between the double occupancy $D=\langle n_\uparrow n_\downarrow \rangle$ from its uncorrelated value of $(\langle n \rangle/2)^2$ provides a third way of studying how far correlations extend away from half filling in the charge response [Fig. 6(c)].  Due to its local nature, this quantity is more sensitive to correlations.  Like the spin-spin correlation function, it shows that they weaken more slowly with doping than the compressibility suggests.  However, unlike the spin-spin correlation function, $(\langle n \rangle/2)^2-D$ shows little particle-hole asymmetry in the doping trend even with next-nearest-neighbor hopping, implying a fundamental difference between the spin and charge responses. 

Figure 6 also highlights the particle-hole symmetry-breaking effect of next-nearest-neighbor hopping on single- and multi-particle quantities.  Although the compressibility shows little doping dependence above $\sim 15\%$ hole or electron doping when $t^\prime=0$, it exhibits noticeable doping dependence on the hole-doped side when $t^\prime=-0.3t$.  The spin correlations are suppressed on the hole-doped side by next-nearest-neighbor hopping, while they are enhanced on the electron-doped side.  Although $D$ is barely impacted by $t^\prime=-0.3t$ because of the double occupancy's local nature, single- and multi-particle quantities that are sensitive to longer-range hopping will display significant particle-hole doping asymmetry.

\section{Conclusions}
In this study, we have investigated spin and charge susceptibilities in the single-band Hubbard model to understand the influence of correlations on the multi-particle response as a Mott insulator evolves into a weakly correlated metal.  The naive expectation based on probes more sensitive to single-particle properties has been for correlations to weaken rapidly with doping such that the cuprates cross-over to a more Fermi-liquid-like behavior for overdoping near the edge  of the superconducting dome.  The compressibility, a quantity associated with the single-particle response, computed using DQMC reflects this behavior as it quickly becomes non-zero and even saturates away from $0\%$ doping.  However, in addition to the dynamical response functions discussed here, two-particle equal-time quantities computed using DQMC also show significant residual correlations to much higher doping levels: the behavior of the spin-spin correlation function counters the naive expectation that the response in the spin channels represents a simple spin flip on top of the charge background, and the difference in the double occupancy from its uncorrelated 
value provides a direct measure of the residual correlations in the system, which extend to relatively high doping levels as concluded from the multi-particle response.  When compared to RPA calculations, the DQMC-derived spin and charge response functions show qualitative differences that persist across large portions of the Brillouin zone and throughout the doping range relevant to the cuprates, especially in the charge channel, attributable to distinctions between the doping dependence of correlation effects at the single- and multi-particle level.  Only when doped to extremely high levels will the two-particle response represent a system in a weakly correlated metallic state.  These conclusions help to elucidate evolution away from the Mott insulating ground state and demonstrate that strong correlations can extend over a larger region of the cuprate phase diagram than has been appreciated previously.

We would like to thank Steve Kivelson, Douglas Scalapino, and Krzysztof Wohlfeld for helpful discussions.  This research was supported by the U.S. Department of Energy (DOE), Office of Basic Energy Sciences, Division of Materials Sciences and Engineering, under Contract No. DE-AC02-76SF00515, SLAC National Accelerator Laboratory (SLAC), Stanford Institute for Materials and Energy Sciences.  Y.F.K. was supported by the Department of Defense (DOD) through the National Defense Science and Engineering Graduate Fellowship (NDSEG) Program and by the National Science Foundation (NSF) Graduate Research Fellowship under Grant No. 1147470.  S.J. acknowledges partial support from The Joint Directed Research and Development (JDRD) program with Oakridge National Laboratory.  The computational work was partially performed at the National Energy Research Scientific Computing Center (NERSC), supported by the U.S. DOE under Contract No. DE-AC02-05CH11231.

\bibliography{RPA_DQMC_Bib}

\end{document}